# Pixie: A Context-Aware Multi-Agent Multi-Modal Large Language Models (LLM) Architecture for Digital Marketing

Vivek Vishwas Vichare
vivek.vichare@pixis.ai

Kirill Dubovikov
kdubovikov@pixis.ai

Vivek Kumar
vivek.kumar@pixis.ai

Sreyas Chaithanya
sreyas.chaithanya@pixis.ai

Saurabh Verma
saurabh.verma@pixis.ai

Humeil Makhija
humeil.makhija@pixis.ai

Akshat Singhal
akshat.singhal@pixis.ai

Mrunal Nasery
mrunal.nasery@pixis.ai

Pixis
All authors are equal contributors

***Abstract*— We present an innovative framework called Pixie, designed to enhance digital marketing operations through a context-aware, multi-agent large language model (LLM) architecture. The framework is versatile, supporting both real-time and offline marketing activities, and leverages multimodal LLM capabilities to increase reliability and performance.**

**Pixie offers a generalized multiagent architecture capable of performing a broad spectrum of digital marketing tasks. This is achieved through cooperation between a primary orchestration agent and task-specific specialized agents. These agents combine different learning techniques, such as zero-shot and few-shot prompt-based learning, with insights derived from marketing data, and proprietary expert marketing strategy guidelines and documentation on our proprietary statistical models. The agents broadly belong to the following categories - Primary Orchestration Agent which coordinates the workflow, understands user queries, and consolidates outputs from all agents; Task-Specific Agents which perform specific tasks such as performance analysis and Reconciliation/Reviewer Agents which evaluate and reconcile outputs from other agents using chain of thought reasoning.**

**We experimented with different agentic workflows, ranging from single agent to multi-agent systems. The framework was evaluated through a combination of human expert assessments (trained marketers) and measuring LLM output variance. The evaluation criteria focused on accuracy, consistency, and the ability to solve domain-specific challenges. The results demonstrated that Pixie's multi-agent framework significantly improves decision quality and operational efficiency in digital marketing workflows.**

**We evaluated the efficacy of the Pixie framework on some of the most frequent digital marketing processes and it can potentially reduce the time spent on tasks such as root cause analysis and reporting, campaign research and planning. The framework is flexible, allowing for the integration of additional specialized agents tailored to specific marketing tasks. This expansion will further enhance the system's capabilities and its ability to adapt to the evolving demands of digital marketing.**

## I. Introduction

Modern digital marketing faces unprecedented challenges due to the explosion of multimodal data and rapidly changing consumer behaviors. Traditional marketing tools are often task-specific, optimized for narrow use cases, such as campaign optimization or performance monitoring. While effective within their domains, they lack the ability to integrate insights, communicate with other systems, or adapt to changing requirements.

Large Language Models (LLMs) have demonstrated significant potential in digital marketing [1], enabling personalized content creation, strategy analysis, and explainable insights generation. However, these models are constrained by challenges such as complex reasoning limitations, numerical errors and a tendency to generate impractical suggestions. LLMs frequently produce hallucinations due to training data restrictions that lack domain-specific or latest information. The rapidly evolving marketing landscape necessitates designing systems which are capable of providing contextually relevant responses, integrating seamlessly across tools, and identifying optimal solutions that align with critical user constraints such as budget limits, brand guidelines, and audience preferences.

Autonomous agents [2], [3], [4] have gained significant traction for their ability to enhance large language models (LLMs) with advanced functionalities, such as context-aware memory [5], multistep planning [6], and capability to learn using strategic tools [7]. These developments aim to create a generalist agentic system that employs a team of specialized agents capable of performing various tasks, such as web browsing, file management, and code execution [8]. This approach is not limited to technology; similar advancements are being observed in sectors like logistics where multi-agent systems can dynamically manage tasks and commitments [9]. In marketing analytics [10] and data science [11], these agents are being utilized to automate the generation of synthetic data and automating machine learning workflows, while in healthcare, they enhance decision-making and improve patient care outcomes [12].

## II. Scope of Work

Motivated by the expert-like potential of autonomous agents, we introduce Pixie, a novel agentic workflow application to the digital marketing domain. Pixie works through co-operation between a primary orchestration agent and task-specific specialized agents. These agents combine different learning techniques, such as zero-shot and few-shot prompt-based learning, with insights derived from marketing data, historical actions, and proprietary expert reasoning documents and documentation on our proprietary statistical models.

Our current work is focused on the most frequent and essential use-cases in performance marketing related to performance analysis and optimization, creative and targeting strategy analysis and insights generation. Our generalized multi-agentic architecture allows developing more granular, hyper-specialized agents for other tasks in future. We have created a platform-independent reasoning framework, aimed at efficiently working in multiple advertising platforms such as Meta and Google.

The different agentic workflows were evaluated for the aforementioned tasks through a combination of human expert assessments (trained performance marketing experts) and measuring LLM-agent output variance. The evaluation criteria focused on accuracy, consistency, and the ability to solve domain-specific challenges.

The next sections are divided into 1. Solution Architecture which explains different agentic workflows we evaluated and in-depth overview of multi-agentic workflow 2. Evaluation Framework and Results which explains our evaluation framework and results for different agentic workflows and individual specialized agents and 3. Conclusions and Future Work.

## III. Solution Architecture

We have experimented with three different types of agentic workflows implemented in increasing order of complexity. We performed this experiment to evaluate and identify optimal architecture for generalized digital marketing agent workflow. All three workflows are implemented in LangGraph and have three essential components:

- Agents - Agents are Large Language Models which are tasked with different roles. We used OpenAI's GPT-4o model because of its larger context window.
- Specialized Task-specific Models Integrated as Tools - We provide specialized LLM agents with access to proprietary statistical recommendation models for various digital marketing tasks such as Performance Analysis, Bid-Budget Optimization, Audience Targeting Refinement and Creative Recommendations. Each task is addressed by an independent, task-specific model integrated as a tool for the corresponding LLM agent.
- Knowledge Infrastructure - Pixis knowledge infrastructure integrates knowledge graph and data warehouse to support contextual data-driven decision-making. The knowledge graph serves as a semantic network, mapping industry vertical relationships, brand hierarchies, and marketing keyword taxonomies. The data warehouse aggregates structured data from major marketing platforms like Meta Ads and Google Ads, capturing campaign performance metrics, historical spending information, and third-party conversion metrics.

The three workflows essentially differ from each other in whether specialized task-specific models are implemented as tools or agents. Furthermore, the workflows differ in how different agents accessed different models as tools. Additionally, we experimented with collaborative and hierarchical agentic structure. Our primary objective behind these experiments was to address whether increased agentic complexity leads to improved performance in terms of accuracy.

### A. *Workflow 1 – Single Agent Architecture*

In single-agent architecture, we deployed a comprehensive digital marketing agent equipped with access to task-specific models. The agent's primary objective is to understand and address user queries related to digital marketing campaign performance analysis and identify measures for improvement.

The workflow involves a direct interaction between the user and the agent, where the user presents specific marketing challenges or seeks optimization recommendations. Upon receiving a query, the agent strategically leverages its tool suite to analyse the problem, extract relevant insights, and formulate data-driven recommendations. This approach enables a holistic and tool-integrated response mechanism, allowing the agent to provide nuanced, context-aware suggestions for enhancing marketing campaign effectiveness.

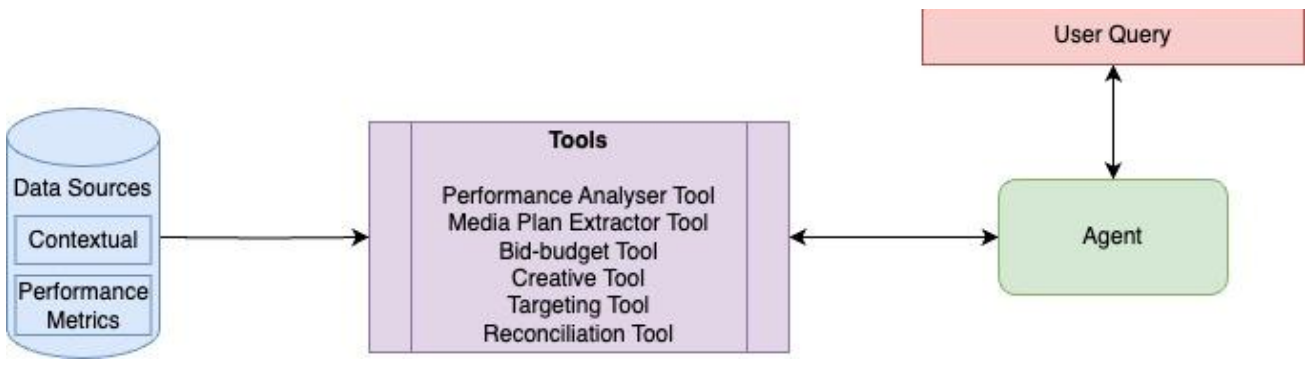

Figure 1: Workflow 1 – Single Agent Architecture

### B. *Workflow 2 – Collaborative Multi-agent architecture*

In this workflow, three specialized agents—Recommender, Analyzer, and Reconciler—are assigned specific roles with access to relevant task specific models. In this architecture, the recommender agent interfaces with the user and then collaborates with the other agents which focus on specific tasks of analyzing recommendations and then reconciling outputs. Unlike the single-agent approach, these agents have a more focused task scope, which reduces hallucinations. However, since the architecture does not involve any hierarchical planning agent for routing tasks, it cannot be scaled for diverse tasks beyond specific task of performance analysis.

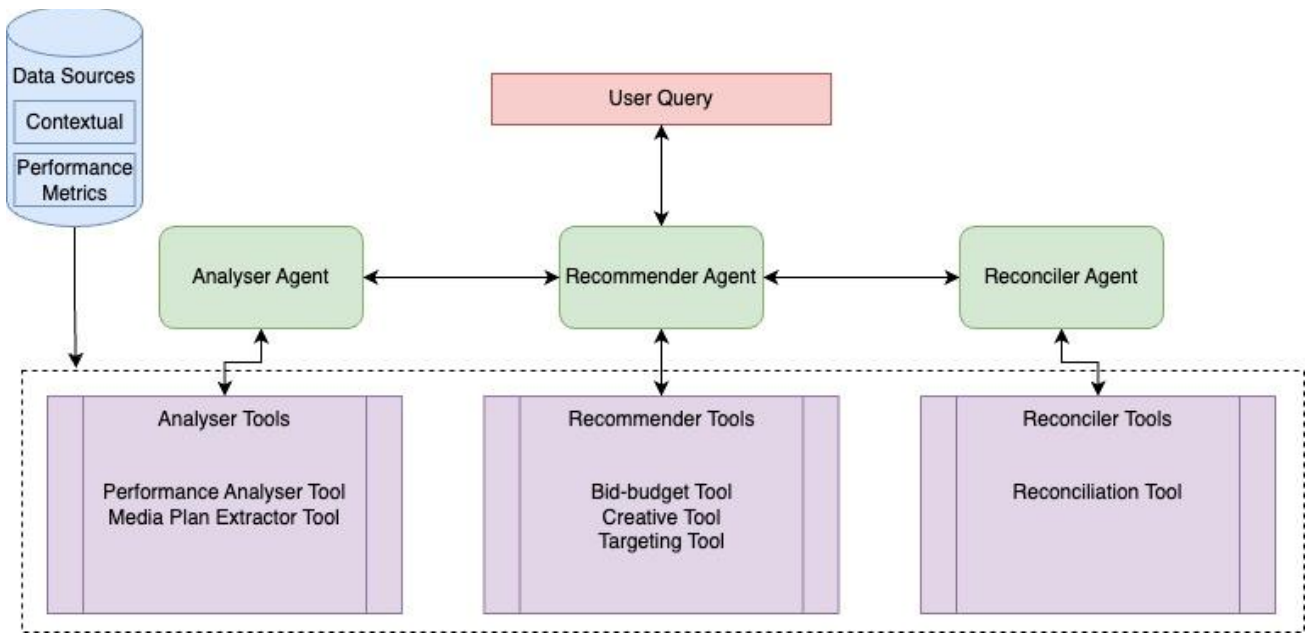

Figure 2: Workflow 2 - Collaborative Multi-agent Architecture

### C. *Workflow 3 – Multi-agent with hierarchical architecture*

We expanded our multi-agent architecture by introducing a hierarchical graph structure with a primary orchestrator agent and seven specialized sub-agents. This approach allowed for increased problem-solving scope and task specialization without compromising output quality and mimicking real world marketing agency workflow which we have studied in our daily work. By introducing a sub-agent with vision capabilities, we transformed the workflow into a multi-modal system. The hierarchical structure enables flexible integration of additional agents, positioning this workflow termed Pixie as a generalized framework for addressing diverse digital marketing tasks and performing deep-dive on each task through follow-up user-interaction.

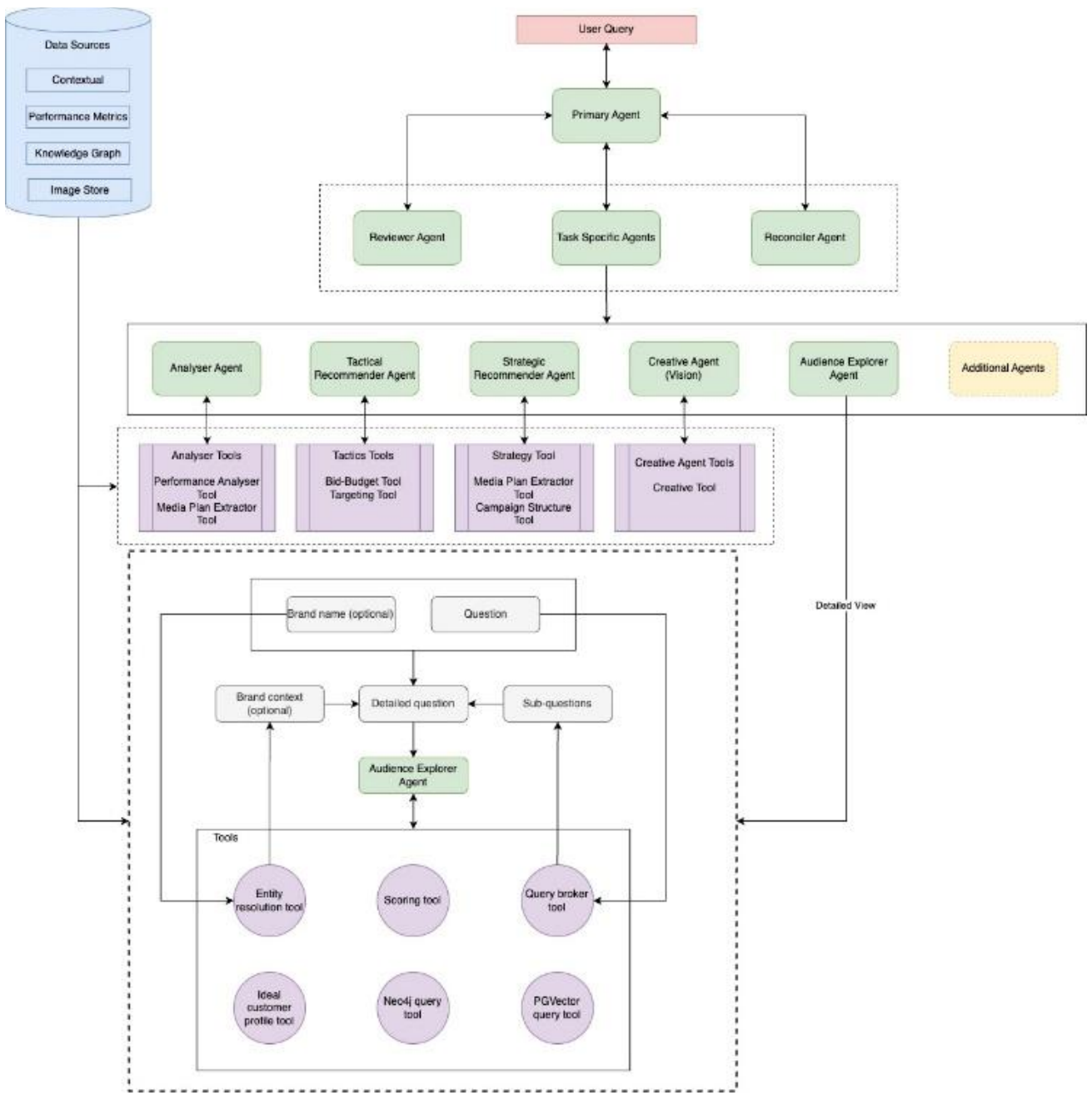

Figure 3: Workflow 3 - Pixie - Multi-agent hierarchical architecture

The individual agents are described in detail as below:

- **Primary Agent** - The primary agent serves as the main interface for user interaction, responsible for comprehending user requirements and strategically delegating tasks to appropriate agents to generate comprehensive recommendations. This agent orchestrates an iterative process of reconciliation and review, coordinating between recommender, reconciler, and reviewer agents before finalizing the report using the writer tool. To enhance task complexity management, we applied zero-shot and few-shot chain-of-thought prompting techniques, enabling the primary agent to decompose complex tasks into smaller, more manageable subtasks, thereby improving overall system performance and output quality.

- **Analyser Agent** - The Analyser Agent, delegated tasks by the Primary Agent, specializes in performance analysis and identifying necessary actions for campaign improvement across tactical (bid/budget and targeting), strategic, creative, and audience exploration dimensions. Utilizing the performance analyser tool and media plan extractor tool, this agent examines current campaign performance and extracts performance goals from strategic documents. Access to arithmetic tools enables precise numerical operations such as calculations and averages. Upon completing its assigned tasks, the Analyser Agent returns control to the Primary Agent, facilitating a streamlined and focused workflow for performance optimization.

- **Recommender Agents** - The Tactical and Strategic Recommender Agents provide targeted recommendations within the multi-agent framework. The Tactical Recommender Agent focuses on generating short-term recommendations to improve campaign performance efficiency, utilizing bid-budget, targeting and creative tools to propose specific bid, budget, and targeting updates. In contrast, the Strategic Recommender Agent develops long-term strategic recommendations by analyzing campaign structure and performance goals using the campaign structure and campaign objective tracker tools. The agents leverage proprietary statistical and machine learning models as tools to perform complex numerical and analytical tasks, ensuring precise and data-driven recommendations across performance analysis, bid-budget and targeting optimization, and creative strategy. Both agents are augmented with insights from marketing data, proprietary expert reasoning documents, and documentation on statistical models which are integrated as tools, enhancing the framework's context-awareness.

- **Audience Explorer Agent:** The Audience Explorer Agent is designed to provide audience segment recommendations for marketing campaigns by intelligently analyzing product preferences and generating customer insights. Leveraging the principles of SELF-DISCOVER [13] and Subgraph RAG [14], the agent systematically breaks down complex use cases into actionable solution steps using a comprehensive suite of tools and advanced knowledge infrastructure. The agent comprises six specialized tools that work synergistically to deliver precise targeting recommendations:

1. **Query Broker Tool**: Employs few-shot prompting to determine optimal execution steps based on knowledge graph and database schemas.

2. **Entity Resolution Tool**: Validates and matches query entities within the Pixis Knowledge Graph, enhancing query accuracy and relevance.

3. **Scoring Tool**: Calculates semantic relevance for entities, enabling prioritized decision-making.

4. **Ideal Customer Profile Tool**: Generates detailed customer profiles using one-shot prompting, characterized by interests, demographics, and behavioral attributes.

5. **Neo4j Query Tool**: A dynamic graph query generator built on GraphCypherQAChain, (Chain for question-answering against a graph by generating

Cypher statements) customized for marketing requirements.

6. **PGVector Query Tool**: A flexible query generation mechanism using PGVector (An implementation of LangChain vectorstore abstraction using postgres database structure) for retrieving contextually aligned entities across database tables.

The agent's intelligence is powered by two databases:

- **Pixis Knowledge Graph:**

  A multi-layered semantic graph featuring:

  1. Brand Layer: Comprehensive brand metadata
  2. Latent Representation Layer: Industries, products, and customer profiles
  3. Platform Targeting Layer: Cross-platform targeting options

- **PGVector Database:**

  A multi-table database optimized for efficient entity searching, particularly handling disconnected data components.

This architecture enables the Audience Explorer Agent to generate precise, actionable marketing insights by seamlessly integrating advanced querying techniques with rich, interconnected data resources.

- **Creative Agent (Vision)** - The Creative Agent is an advanced vision-enabled analytical system designed to provide comprehensive creative assessment for advertising campaigns. Utilizing a Vision Large Language Model (GPT-4 Vision), the agent conducts multi-dimensional analysis of ad creatives through intricate evaluation frameworks. The system performs granular assessment of both textual and visual elements, examining text attributes such as occlusion characteristics, clutter density, color and size variations, and cross-relevance between multiple text elements. Simultaneously, it analyzes product visual components across critical dimensions including color composition, thematic coherence, product size and spatial prominence, attention generation metrics, and background relevance. An integrated creative performance tool enables the agent to generate actionable recommendations, specifically determining whether to activate or deactivate specific creative assets based on comprehensive visual analysis. Upon completing its analytical task, the Creative Agent autonomously returns control to the Primary agent, ensuring seamless workflow integration and intelligent creative optimization.

- **Reconciler and Reviewer Agents -** The recommendation generation process incorporates two critical quality assurance mechanisms: the Reconciler and Reviewer Agents. These agents serve as rigorous checkers that critique and refine recommendation reports before final user presentation. The Reconciler Agent focuses on identifying and resolving potential inconsistencies or gaps in recommendations generated by different specialized agents. When conflicting or missing recommendations are detected, this agent instructs the Primary agent to regenerate recommendations with updated guidance, ensuring comprehensive and coherent outputs. Similarly, the Reviewer Agent evaluates reports for critical attributes such as clarity, specificity, and explainability. If the reports fall short of these standards, the agent triggers a refinement process wherein the Primary agent re-delegates tasks to the appropriate recommender agents for improvement. To prevent infinite iteration, the review process is strategically limited to three feedback rounds. This systematic approach of iterative refinement ensures that the final recommendations meet high-quality standards, providing users with precise, well-structured, and actionable insights derived from a collaborative and self-critical multi-agent system.

## IV. Experiments and Evaluation Framework

### A. *Workflow Evaluation Methodology*

The evaluation of three agentic workflows and the individual specialized agents for the multi-agent workflow were made with the help of expert digital performance marketers. LLM response variability was evaluated for consistency and repeatability through multiple sampling of the same query. Expert human marketers interacted with the agentic workflows through a series of questions addressing tasks related to performance analysis and efficiency across different dimensions. The evaluation was repeated for a total of four diverse marketing campaigns across geography, platform and spends. The experts then evaluated the quality of the responses on the basis of various dimensions listed below on a scale of (0,1,2) - 0 indicating the worst acceptance, 1 being partial and 2 being absolute agreement with the response [15]. The different metrics of evaluation are as follows:

- **Conciseness**: This metric evaluates how effectively the response conveys its message without unnecessary repetition or extraneous details.
- **Explainability**: This metric evaluates how clearly the response articulates its reasoning, ensuring that the rationale behind the answer is transparent and understandable.
- **Coherence**: This metric assesses the logical and grammatical consistency of the response, ensuring that all parts of the text flow smoothly and relate meaningfully to one another.
- **Reasoning accuracy**: This metric evaluates the logical correctness of the thought process behind the response.
- **Recall**: This metric measures the response’s ability to retrieve and include all relevant information required to address the query comprehensively.
- **Numerical accuracy**: This metric focuses on the correctness of numerical information in the response, including values, calculations, or quantitative comparisons.
- **Factual accuracy**: This metric assesses the factual correctness of the response, focusing on whether the information provided aligns with verified facts from the ground truth and the available knowledge base.

### B. *Individual Agent Evaluation Methodology*

For Pixie workflow, we employed a holistic evaluation framework to assess the performance of agents within the hierarchical multi-agent workflow. The framework consisted of systematically measuring the quality of individual agent's response using a similar framework as used for evaluating different agentic workflows.

| Workflow name | User Question | Average Latency (in seconds) | Standard Deviation Latency (in seconds) | Average Cost (in $) | Std dev Cost (in $) |
|---|---|---|---|---|---|
| Single Agent | Query 1 - Can you provide tactical recommendations for this campaign/adsets? | 187.80 | 133.92 | 0.10 | 0.04 |
| | Query 2 - Can you provide strategic recommendations for this campaign/adsets? | 37.39 | 8.15 | 0.03 | 0.00 |
| | Query 3 - Can you summarize the performance of this campaign/adsets? | 43.78 | 10.03 | 0.03 | 0.01 |
| | Query 4 - Can you recommend the optimal actions for this campaign/adsets based on its performance? | 182.67 | 144.41 | 0.09 | 0.05 |
| Collaborative Multi-Agent | Query 1 - Can you provide tactical recommendations for this campaign/adsets? | 214.35 | 135.04 | 0.20 | 0.08 |
| | Query 2 - Can you provide strategic recommendations for this campaign/adsets? | 212.13 | 9.31 | 0.08 | 0.03 |
| | Query 3 - Can you summarize the performance of this campaign/adsets? | 242.36 | 148.58 | 0.12 | 0.10 |
| | Query 4 - Can you recommend the optimal actions for this campaign/adsets based on its performance? | 213.79 | 138.05 | 0.13 | 0.02 |
| Hierarchical Multi-Agent | Query 1 - Can you provide tactical recommendations for this campaign/adsets? | 112.17 | 34.06 | 0.16 | 0.03 |
| | Query 2 - Can you provide strategic recommendations for this campaign/adsets? | 52.94 | 11.67 | 0.10 | 0.01 |
| | Query 3 - Can you summarize the performance of this campaign/adsets? | 38.83 | 17.28 | 0.07 | 0.10 |
| | Query 4 - Can you recommend the optimal actions for this campaign/adsets based on its performance? | 1004.60 | 1039.92 | 0.55 | 0.17 |

Table 1 - Latency and Cost Metrics for different agentic workflows

### C. *Agent variance evaluation methodology*

Further, we evaluated consistency and reliability of each agent response through repeated executions. We executed the LLM workflow 10 times for each of four diverse marketing campaigns, generating 10 responses per query per agent. To analyze variance, we computed text distance metrics including factual accuracy, response structure, and entity identification across these responses. We calculated the mean distance between responses for each unique query-campaign pair for every agent. Then, we created variance distributions using box plots. The evaluation process aimed to quantify system stability by examining how closely variance values approached zero, with consistently low variance indicating high reliability. By rigorously testing agents across multiple dimensions and repeated trials, we sought to validate the robustness of the multi-agent workflow and its ability to generate consistent, high-quality recommendations across different marketing contexts. We performed this analysis for both individual agent responses as well as every workflow's response for different queries.

## V. Results

Table 1 summarizes cost and latency metrics for different agentic workflows. The latency metric includes response time of the agent as well as computation time for the tools associated with each agent. As the complexity of the workflow architecture increases, the cost and latency metrics increase. Table 2 summarizes the human expert evaluation results across different metrics for different agentic workflows. The increase in cost and latency metrics are offset by an improvement in the response quality of the workflow.

Thus, our experiments demonstrate the reliability of the hierarchical multi-agentic workflow as opposed to the agentic workflows in single-agent and collaborative multi-agent frameworks. This further concretises the hypothesis that adding more specialised agents will help improve the accuracy of metrics across the board. Better scores for explainability and reasoning for Pixie highlight that as we convert individual statistical tools into agents, they become better at explaining the 'black box' outcomes of statistical models. Although based on limited but comprehensive testing, these preliminary results will need to be further corroborated with more data collection.

| Workflow | Conciseness | Explainability | Coherence | Reasoning | Recall | Numerical Accuracy | Factual Accuracy |
|---|---|---|---|---|---|---|---|
| Single Agent | 1.73 (0.47) | 1.10 (0.70) | 1.36 (0.50) | 1.36 (0.50) | 1.36 (0.50) | 1.82 (0.40) | 1.91 (0.30) |
| Collaborative | 1.75 (0.45) | 1.08 (0.79) | 1.33 (0.65) | 0.67 (0.49) | 0.92 (0.51) | 1.58 (0.79) | 1.75 (0.62) |
| Multi-Agent | | | | | | | |
| Hierarchical Multi-Agent | 1.63 (0.50) | 1.75 (0.45) | 1.56 (0.51) | 1.63 (0.50) | 1.69 (0.48) | 2 (0.00) | 1.88 (0.34) |

Table 2: Accuracy metrics (Mean / Standard Deviation) for agentic workflows

Our experimental evaluation revealed better performance across different workflows, particularly in numerical and factual accuracy, which significantly outperformed conventional expectations for LLMs [16]. The near-perfect score can be attributed to the integration of task-specific agents with specialized statistical and numerical tools that enable deterministic calculations. While the coherence score was comparatively lower, this was primarily cosmetic feedback from human experts aimed at refining the summary's structural presentation, which does not compromise the underlying reasoning or explainability. However, the relative inconclusiveness between single-agent and collaborative multi-agent workflows warrants further investigation to definitively establish the comparative advantages of each approach.

| **Agent Type** | **Conciseness** | **Explainability** | **Coherence** | **Reasoning** | **Recall** | **Numerical Accuracy** | **Factual Accuracy** |
|---|---|---|---|---|---|---|---|
| Primary Agent | 2 (0) | 2 (0) | 2 (0) | 2 (0) | 2 (0) | 2 (0) | 2 (0) |
| Creative Agent (Vision) | 2 (0) | 1 (0) | 2 (0) | 2 (0) | 2 (0) | 2 (0) | 2 (0) |
| Audience Explorer Agent | 2 (0) | 2 (0) | 2 (0) | 2 (0) | 1.75 (0.50) | 2 (0) | 2 (0) |
| Analyser Agent | 2 (0) | 1.33 (0.58) | 2 (0) | 2 (0) | 1.67 (0.58) | 2 (0) | 2 (0) |
| Strategic Recommender Agent | 2 (0) | 1.67 (0.58) | 2 (0) | 2 (0) | 2 (0) | 2 (0) | 1.67 (0.58) |
| Tactical Recommender Agent | 1.5 (0.58) | 2 (0) | 2 (0) | 2 (0) | 2 (0) | 2 (0) | 2 (0) |

Table 3 - Accuracy metrics (Mean / Standard Deviation) for different agents in Pixie workflow

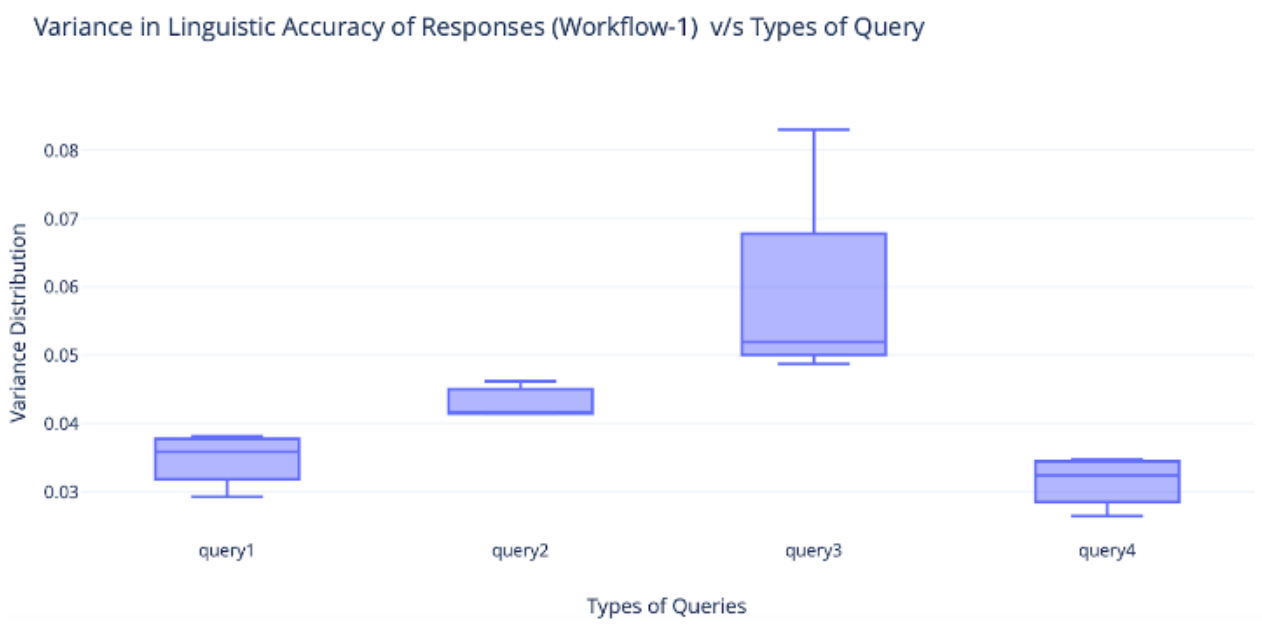

Figure 4 – Variance in responses of single agent workflow

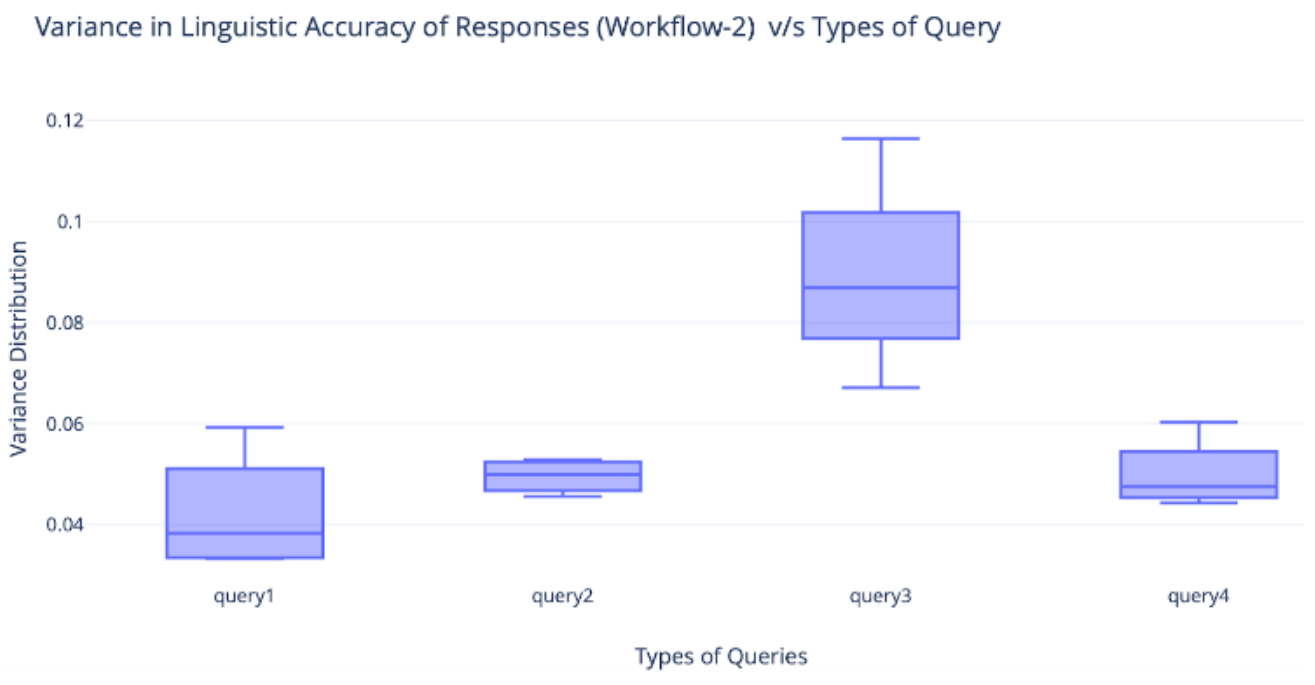

Figure 5 – Variance in responses of multi agent collaborative workflow

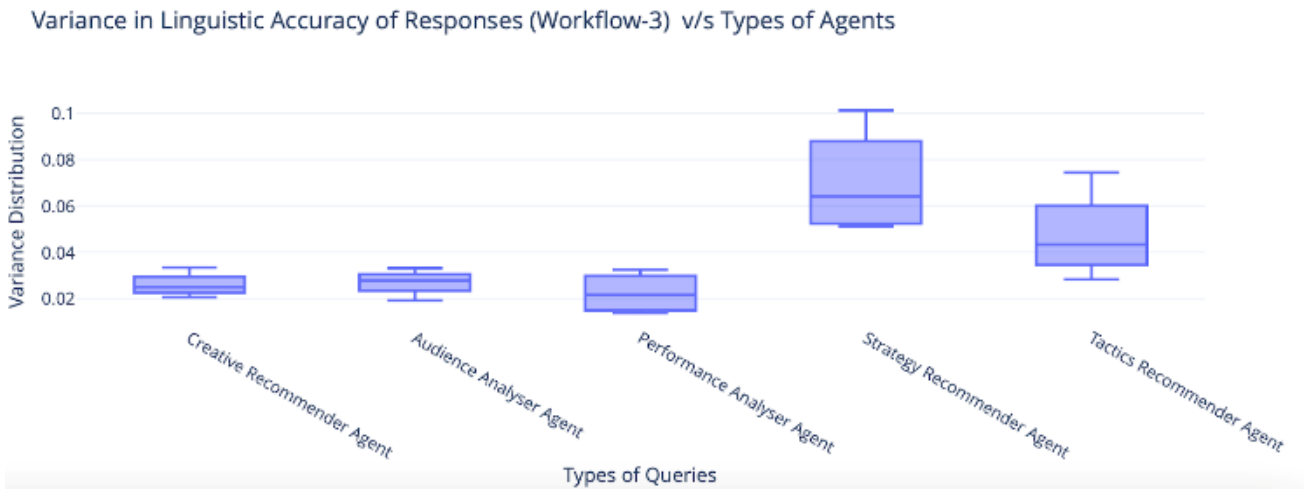

Figure 6 – Variance in responses of different agents in Pixie workflow

Table 3 summarizes results of human expert evaluations for individual agents in the hierarchical multi-agent workflow. Figures 4 and 5 summarize the variance box-plots for single agent and collaborative multi-agent workflows. Figures 6 summarize the variance box-plots for all the individual agents for Pixie. The results demonstrate robust performance of agents across all the dimensions. High performance scores with low variance for performance analyzer, creative and audience explorer agents demonstrate their consistency and stability. Agents which rely on the combination of their tools and general intelligence such as Strategy Recommender agent has high performance scores and but slightly high variance suggest that these agents need further fine-tuning in-order to maintain their consistency.

Overall, higher accuracy scores on some of the most frequent digital marketing processes demonstrate that LLM based multi-agentic workflows can provide a huge efficiency boost to digital marketer’s tasks which take up to 50%- 60% of their daily bandwidth.

## VI. Conclusions and Future Work

Pixie framework presents a novel approach to digital marketing operations, and our research opens up several promising avenues for future investigation and development. This section outlines potential research directions that can build upon our current work and further advance the field of multi-agent language model systems in digital marketing.

Our current framework demonstrated significant potential in handling diverse marketing tasks, but there remains substantial room for expanding agent specialization. Future research could focus on developing more granular, hyper-specialized agents targeting specific marketing sub-domains such as media-planning, competitor intelligence, marketing-mix optimization etc. By creating increasingly sophisticated domain-specific agents, we can enhance the framework's ability to provide nuanced, context-aware marketing insights. The reconciliation agent represents a critical component of our multi-agent architecture. Future work could investigate more sophisticated reasoning and validation techniques, such as implementing more complex chain-of-thought reasoning algorithms. Our preliminary results suggest significant time savings in marketing operations. Our next steps include developing a more robust scalable architecture and reducing computation overhead in multi-agent interactions. Further, we propose several experimental directions to validate and extend our framework through comparative studies with alternative multi-agent architectures and stress testing the framework under more diverse marketing scenarios.

The Pixie framework represents a significant step towards more intelligent, adaptive digital marketing systems. By continuing to refine our multi-agent approach, we can develop increasingly sophisticated tools that not only automate marketing processes but also provide strategic, context-aware insights.